# Critical slowing down and attractive manifold: a mechanism for dynamic robustness in yeast cell-cycle process

Yao Zhao[1,2,*], Dedi Wang[1,2,*], Zhiwen Zhang[1,2], Ying Lu[3], Xiaojing Yang[2], Qi Ouyang[1,2], Chao Tang[1,2], and Fangting Li[1,2,†]

1 School of Physics, Peking University, Beijing 100871, China
2 Center for Quantitative Biology, Peking University, Beijing 100871, China
3 Department of Systems Biology, Harvard Medical School, Boston, MA 02115, USA

[†] To whom correspondence should be addressed. lft@pku.edu.cn

[*] These authors contributed equally to this work.

## Abstract

The biological processes that execute complex multiple functions, such as cell cycle, must ensure the order of sequential events and keep the dynamic robustness against various fluctuations. Here, we examine the dynamic mechanism and the fundamental structure to achieve these properties in the cell-cycle process of budding yeast *Saccharomyces cerevisiae*. We show that the budding yeast cell-cycle process behaves like an excitable system containing three well-coupled saddle-node bifurcations to execute DNA replication and mitosis events. The yeast cell-cycle regulatory network can be separated into G1/S phase module, early M module and late M phase module, where the positive feedbacks in each module and the interactions among the modules play important role. If the cell-cycle process operates near the critical points of the saddle-node bifurcations, there is a critical slowing down or ghost effect. This can provide the cell-cycle process with a sufficient duration for each event and an attractive manifold for the state checking of the completion of DNA replication and mitosis; moreover, the fluctuation in the early module/event is forbidden to transmit to the latter module/event. Our results suggest both a fundamental structure of cell-cycle regulatory network and a hint for the evolution of eukaryotic cell-cycle processes, from the dynamic checking mechanism to the molecule checkpoint pathway.

## Introduction

The eukaryotic cells replicate themselves through the cell-cycle process with a well-programmed sequence of events, including DNA synthesis in S phase, chromosome separation in mitosis phase (M phase), and the cytokinesis event (cell division). G1 phase is the gap after the cell division and before the next DNA replication, and G2 phase is the gap after DNA replication and before nuclear division. To ensure the stability of genetic information, the cell-cycle processes that are controlled by the cell-cycle regulatory network should be stable and robust to the various environmental conditions and the noise inside the cells (McAdams & Arkin, 1999). In addition, the eukaryotic cell-cycle processes are shown to be composed by a series of biochemical switches that control the DNA replication and mitosis events in sequential order (Alberts, Johnson, Lewis, & Ma, 2015). The checkpoint pathways in cell-cycle process help to ensure the ordering progression of cell-cycle and the completion of early events before later events begin, which are indispensable for normal cell-cycle in monitor processes, such as DNA replication and chromosome alignment (Bartek, Lukas, & Lukas, 2004; Hartwell & Weinert, 1989; Nurse, 2000). The checkpoint pathway is usually consisted of sensor proteins, signal transduction kinases, and effect proteins.

In recent years, the quantitatively biological researches in cell-cycle process, especially in the single-celled model organism budding yeast Saccharomyces cerevisiae, have increased our understanding of the regulatory mechanism in cell-cycle process (Verdugo, Vinod, Tyson, & Novak, 2013). The cell-cycle process is controlled by a series of genetic switches, including entrance of S phase (Bean, Siggia, & Cross, 2006; Liu et al., 2015; Yang, Lau, Sevim, & Tang, 2013), entrance of M phase (Gardner, Putnam, & Weinert, 1999), and the metaphase/anaphase (M/A) transition (Lu & Cross, 2010). Modeling the cell-cycle processes, especially the positive feedback and nonlinear interactions in the cell-cycle regulatory network, together with the quantitative experimental results, have highlighted the regulatory mechanism in cell-cycle process from a more quantitative and systemic perspective (J. E. Ferrell, Jr., Tsai, & Yang, 2011).

In this paper, we utilize nonlinear dynamic analysis to study the global dynamic properties in the cell-cycle process of budding yeast. We propose to reveal the dynamic mechanism and the fundamental structure of the regulatory networks to ensure the robustness of cell-cycle process (as a time trajectory) against the fluctuations and noise. Our results show that the yeast cell-cycle process and G1 state are dynamically and structurally stable and robust against the changing of both initial states and kinetic parameters. The nonlinear analysis and parameter sensitivity analysis suggest that the yeast cell-cycle regulatory network can be separated into G1/S module, early M module and late M module, where the positive feedbacks in each module and the interactions among the modules play important role in governing the yeast cell-cycle process. The positive feedbacks in each module cause the saddle-node bifurcations that provide the genetic switches for the key state transitions in cell-cycle process, while the balance among modules ensures the sequential DNA replication and mitosis events. The failure of the balance should result in new attractors in S phase, early M phase, or limit cycle. Furthermore, if the cell-cycle process operates near the critical points of the saddle-node bifurcations, there exists a critical slowing down or ghost effect, and the cell-cycle process behaves as a robust dynamical trajectory.

# Results

In this work, we first constructed an auto-evolving simplified cell-cycle model to simulate the cell-cycle process in the wild type and mutant stains in budding yeast. We compared and modified the simulation results with the yeast experimental data to obtain a set of kinetic parameters to depict wild type cell-cycle process. Then, we analyzed the dynamic stability and robustness of the cell-cycle model in both the state and parameter space. We developed a density map method in the state space to show that the cell-cycle process or trajectory is a global attracting trajectory. The bifurcation analysis results in the parameter space revealed the important interactions of our model and suggested the fundamental structure governing the yeast cell-cycle process. Finally, we analyzed an if-then cell-cycle model to depict checkpoint pathways in cell-cycle process, and discussed the possible evolution steps of the checkpoint pathway in eukaryotic cell-cycle processes.

**Modeling the yeast cell-cycle process: the regulatory network and the ordinary differential equations (ODEs)**

The schematic regulatory network governing the cell-cycle process in budding yeast is shown in Figure 1, where we marked the G1/S phase, early M and late M phase modules that control the DNA replication, mitosis and cytokinesis events respectively. The positive feedbacks can be found in each module, such as: Cln2 and SBF, Clb5 and MBF in G1/S phase module, Clb2 and Mcm1 in early M phase module, Cdc14 in late M module; these positive feedback loops together with negative feedback loops and other interactions compose the regulatory network governing the yeast cell-cycle process. More details see SI.1.

Aiming to analyze the global dynamic property in yeast cell-cycle process, and based on the above cell-cycle network, we constructed a set of simplified auto-evolving ordinary differential equations (ODEs) to describe the cell-cycle process in budding yeast. The equations and parameters are based on the previous model studies (Katherine C. Chen et al., 2004; K. C. Chen et al., 2000; Frederick R. Cross, 2003; F. R. Cross, Archambault, Miller, & Klovstad, 2002; Li, Long, Lu, Ouyang, & Tang, 2004). The concentrations of key regulators and their complex in Fig.1 are treated as variables. Our cell-cycle model has 22 independent variables of cyclins, inhibitors, degraders and transcription factors and 94 kinetic parameters. The ODEs and details can be found in SI.2 and the Supplementary Table I, II and III.

In our simplified auto-evolving ODEs model, we make the following assumptions.

First, only G1 cyclin Cln3 is triggered by cell mass, while other cyclins Cln2, Clb5 and Clb2 are driven by interdependent transcriptions. So the cell mass is decoupled from [Cln2], [Clb5]$_T$ and [Clb2]$_T$ in our model. Moreover, the protein Whi5, an inhibitor of SBF/MBF, is ignored in our model, letting Cln3 activate SBF and MBF directly.

Second, the DNA replication checkpoint and Spindle checkpoint are taken into account, while the role of DNA damage checkpoint is ignored. Variable [DNA] is introduced to represent the DNA replication process, which is activated by a continuous Hill function of [Clb5] with kinetic coefficient $n_{DNA}$. Then the activated [DNA] tiggers the activtion of Mcm1/SFF (M phase transcription factor) through kinetic coefficient $\varepsilon_{mcm,DNA}$, while we set $\varepsilon_{mcm,DNA} = 0$ to represent the on state of DNA replication checkpoint. This is a simplified checkpoint mechanism, and is different from the "if-then" rules (if the DNA replication event is finished, then the early M phase transcription factor Mcm1/SFF is activated). Similarly, [SP], activated by [Clb2], is introduced to represents the spindle assemble and separation process; a high level of [SP] will trigger the forming of Cdc20/APC-P complex through kinetic coefficient $k_{a,20}$ in our model, resulting in the metaphase/anaphase transition. If we set $k_{a,20} = 0$, this case represents the on state of Spindle checkpoint.

Third, the regulatory pathway between Cdc20/APC-P and Cdc14 is simplified, and the positive feedback loop of Cdc14 is added for recent experiment result (Holt, Krutchinsky, & Morgan, 2008).

**Simulations of the yeast cell-cycle process in the wild type and mutation yeast strains**

Based on the previous models (Katherine C. Chen et al., 2004; K. C. Chen et al., 2000) and our knowledge about the yeast cell-cycle (detail in SI 2.2), as well as the durations in G1, S, G2 and M phases, we obtained a set of parameters (Supplementary Table III) to depict the wild type yeast cell-cycle process, noted as the wild type cell-cycle parameters. Utilizing the ODEs in Supplementary Table I together with these parameters, we simulated the yeast cell-cycle process in the wild type yeast strains. Starting from the excited G1 state (initial condition) listed in the Supplementary Table X, we obtained the temporal evolution of key regulator concentrations in the wild type cell-cycle process showed in Fig.2. Because the Cln3 signal decays gradually, we simulated one "cycle" of cell-cycle process.

These results were carefully compared with the microarray data of Mat-alpha budding yeast cell (Spellman et al., 1998) to ensure the right time order of cell-cycle events, as well as the abundance of cyclins and Sic1 with the measurement of Cross's group (F. R. Cross et al., 2002) to ensure the right concentration range.

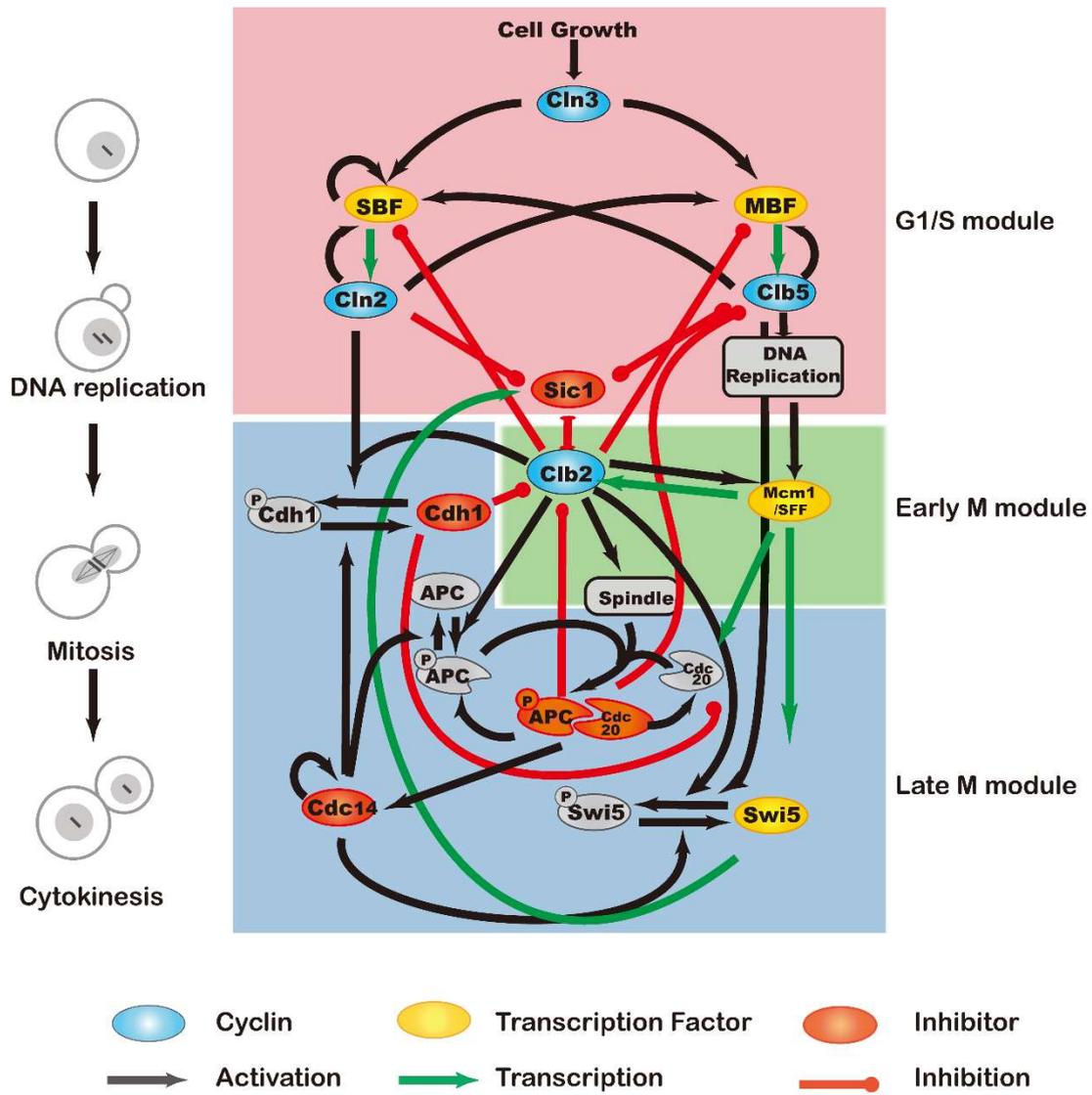

Fig. 1. The key cell-cycle events and the schematic regulatory network in budding yeast cell-cycle process, highlighting the main regulatory modules. The yeast cell-cycle process consists of several key events, including the DNA replication in S phase, mitosis in M phase, and cytokinesis. These sequential multiple events are governed by the relevant G1/S phase module, early M module and late M phase modules, where each module contains the positive feedback loop. The DNA checkpoint is represented by the interaction from "DNA Replication" node to Mcm1/SFF, while spindle checkpoint by the interaction from "Spindle" node to Cdc20/APC-P complexes.

A few of the cell-cycle processes in the key regulator genes knock-out mutants were simulated with our ODEs with the wild type parameters except for setting relative parameters to zero, such as *cln1Δcln2Δcln3Δ* mutant ($k_{s,n2} = k'_{s,n2} = 0$), *clb1Δclb2Δ* mutant ($k_{s,b2} = k'_{s,b2} = 0$), *clb5Δclb6Δ* mutant ($k_{s,b5} = k'_{s,b5} = 0$), *cln1Δcln2Δclb5Δ clb6Δ* mutant ($k_{s,n2} = k'_{s,n2} = k_{s,b5} = k'_{s,b5} = 0$), and *cdc20Δ* mutant ($k_{s,20} = k'_{s,20} = 0$) (Supplementary Table XI). Our simulation results are consistent with experimental observations on these mutants.

The above simulation results of cell-cycle processes in wild type and mutation yeast strains supported the above assumptions in our model; the assumption is that: only G1 cyclin Cln3 is triggered by the cell mass or cell size, other cyclins Cln2, Clb5, and Clb2 are driven by interdependent transcriptions not by the cell mass. So our yeast cell-cycle model is an auto-evolving event transmission model rather than a cell mass central control model.

More details about our ODEs and wild type parameters are presented in SI.2. We noted the above ODEs with the wild type (WT) parameters as the wild type (WT) yeast cell-cycle model.

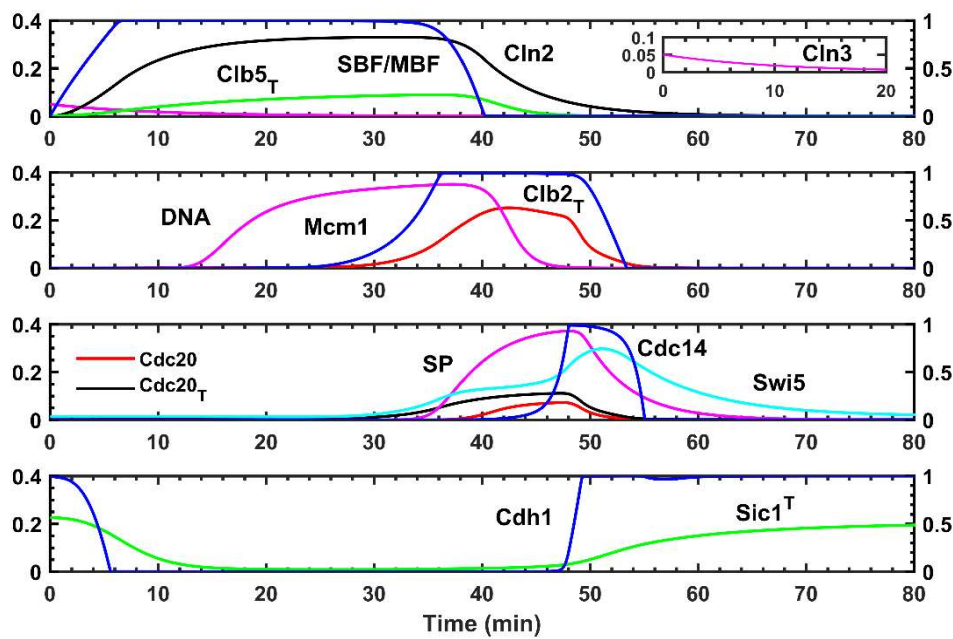

Fig. 2 The simulation results of the wild type yeast cell-cycle model from the excited G1 state as the initial states. The $[X]_T$ represents the total concentration of $X$ variable, while $[Sic1]^T$ represents the total concentration of un-phosphorylated Sic1, [Cdc20] denotes Cdc20/APC-P complex concentration, and [Cdh1] denotes the concentration of Cdh1/APC. The blue line in each panel has a vertical scale

from 0 to 1 (right Y-axis), while others are in the range from 0 to 0.4 (left Y-axis).

**The global attractor G1 and the global attractive cell-cycle trajectory in the wild type cell-cycle model**

We have shown that our model catches the general features of the cell-cycle process in the previous subsection. The next step is to study its global dynamic properties. In a budding yeast cell, it is thought that the resting G1 state (G1 attractor) and the cell-cycle process should both be stable and robust against state and kinetic parameter fluctuations. This requires that the resting G1 state is a global attractor; and the cell-cycle trajectory is a converging trajectory in state space. We demonstrated these dynamic properties with a discrete Boolean network model (Li et al., 2004); here we investigate these properties using the continuous ODEs model.

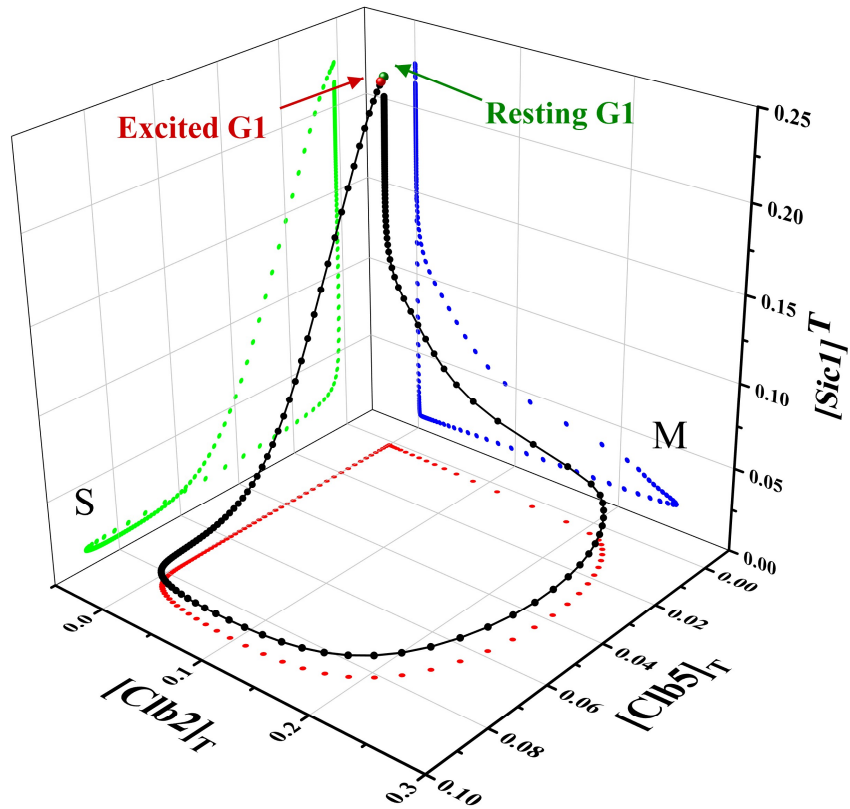

Fig. 3 The dynamic trajectory of wild type yeast cell-cycle process in the 3-demensional state space, $[Clb5]_T$-$[Clb2]_T$-$[Sic1]^T$ space, which correspond to the key regulator in S phase, early M phase and late M phase respectively.

To illustrate the cell-cycle trajectory by our ODEs with different initial states and various parameter sets, we choose three key variables in different cell-cycle phases; they are $[Clb5]_T$ in S phase, $[Clb2]_T$ in early M phase and $[Sic1]^T$ in late M phase. First, the wild type cell-cycle trajectory in Fig.2 is plotted in the $[Clb5]_T$-$[Clb2]_T$-$[Sic1]^T$ state space (Fig.3). Then we utilize our wild type model but starting from random initial states ($10^6$). The initial states are selected from 0 to 1.2 times maximums of each protein concentration in Fig.2 by Latin hypercube sampling, with constraints that the total concentration of a certain protein should be larger than part of it (such as the Sic1, Clb5/Sic1, Clb2/Sic1, Clb5/Sic1$_P$, Clb2/Sic1 $_P$). We trace each trajectory and project it in the 2-dimensional state space, using $[Clb5]_T$-$[Sic1]_T$, $[Clb5]_T$-$[Clb2]_T$ and $[Clb2]_T$-$[Sic1]^T$ as variables, where the 2-demension space is divided into $100 \times 100$ meshes, and we calculate in each mesh the number of trajectories (left panel of Fig.5) or mean transition time (right panel of Fig.5). In Fig. 5, a single stable fixed point (resting G1 state) is found, and no limit cycle is observed. This result suggests that the resting G1 state is the only global attractor of the system, which is consistent with our previous findings [Li et al. 2004]. Furthermore, in the left panel of Fig.5, most of the trajectories converge to the wild type cell-cycle trajectory (black dotted line) and form a ridge, whose density is 2~3 orders of magnitude higher than that out of it. In the right panel of Fig.4, we find that besides the rest G1 attractor, most trajectories cost more time in the area of S phase and M phase.

To show a more comprehensive picture of this attracting pathway, we plot in Fig.5a a sketch of the deviation of random trajectories from the WT process at different WT trajectory distance. We synchronized all the random trajectories at the end of WT cell-cycle process (Sync point), and calculated the Euclidean distance between the corresponding points on two trajectories in phase space (all the variables are first normalized by their maximum concentrations in Fig.2). The distribution of the trajectories is shown in histograms along three different profiles (I, II, III). This result shows that the wild type cell-cycle trajectory doesn't perform attractiveness invariably; instead, it attracts the random trajectories at S phase state and early M phase state, where this S phase state is near to S phase attractor with DNA checkpoint on and early M phase state is near to the M phase attractor with spindle checkpoint on. Together, these show that the cell-cycle trajectory is a global robust attractive trajectory in the state space.

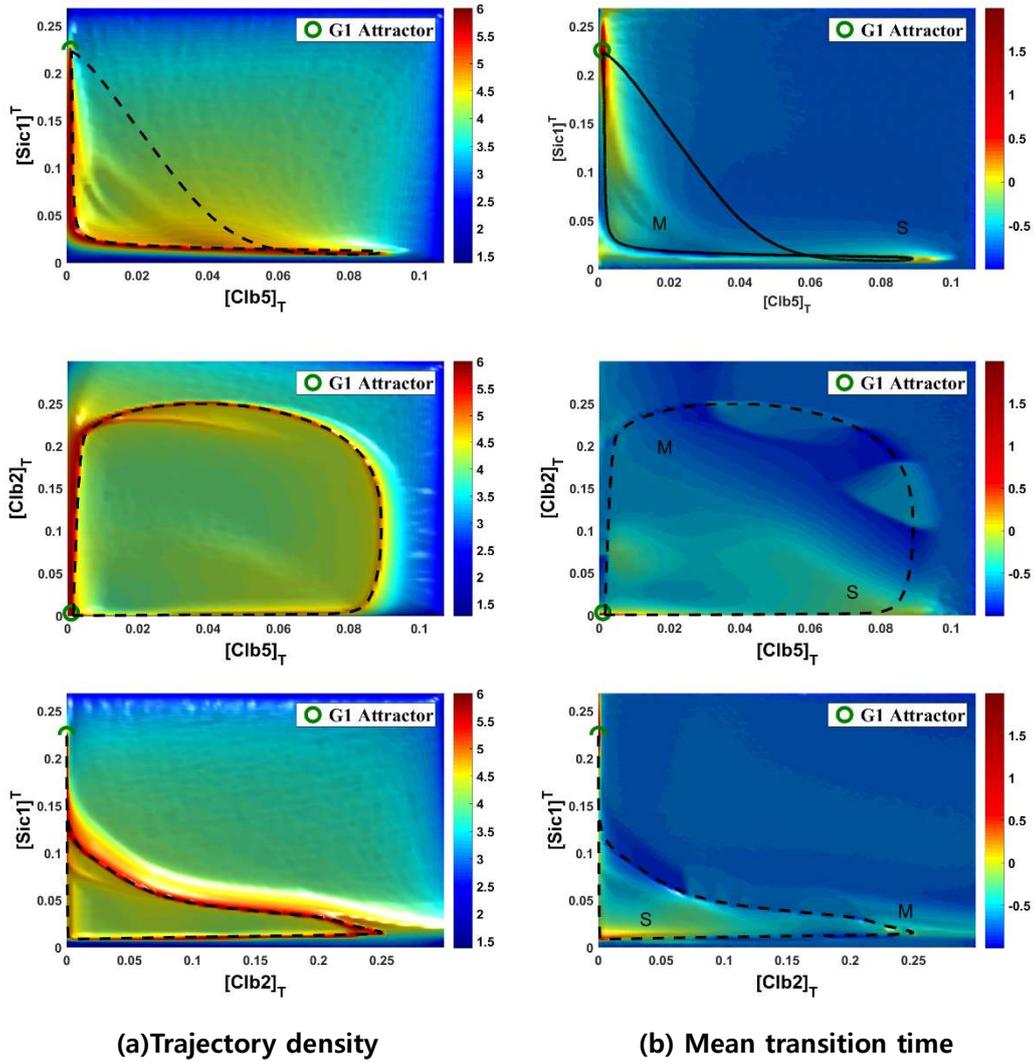

(a) Trajectory density     (b) Mean transition time

Fig.4 Density maps of cell-cycle trajectories of the wild type model from $10^6$ random initial states. The phase space is divided into $100 \times 100$ meshes, and in each mesh the number of trajectories (left panel) or mean transition time (unit: min; right panel) is calculated. The color bar has been set to a logarithmic scale. The black dotted line represents the wild type cell-cycle trajectory.

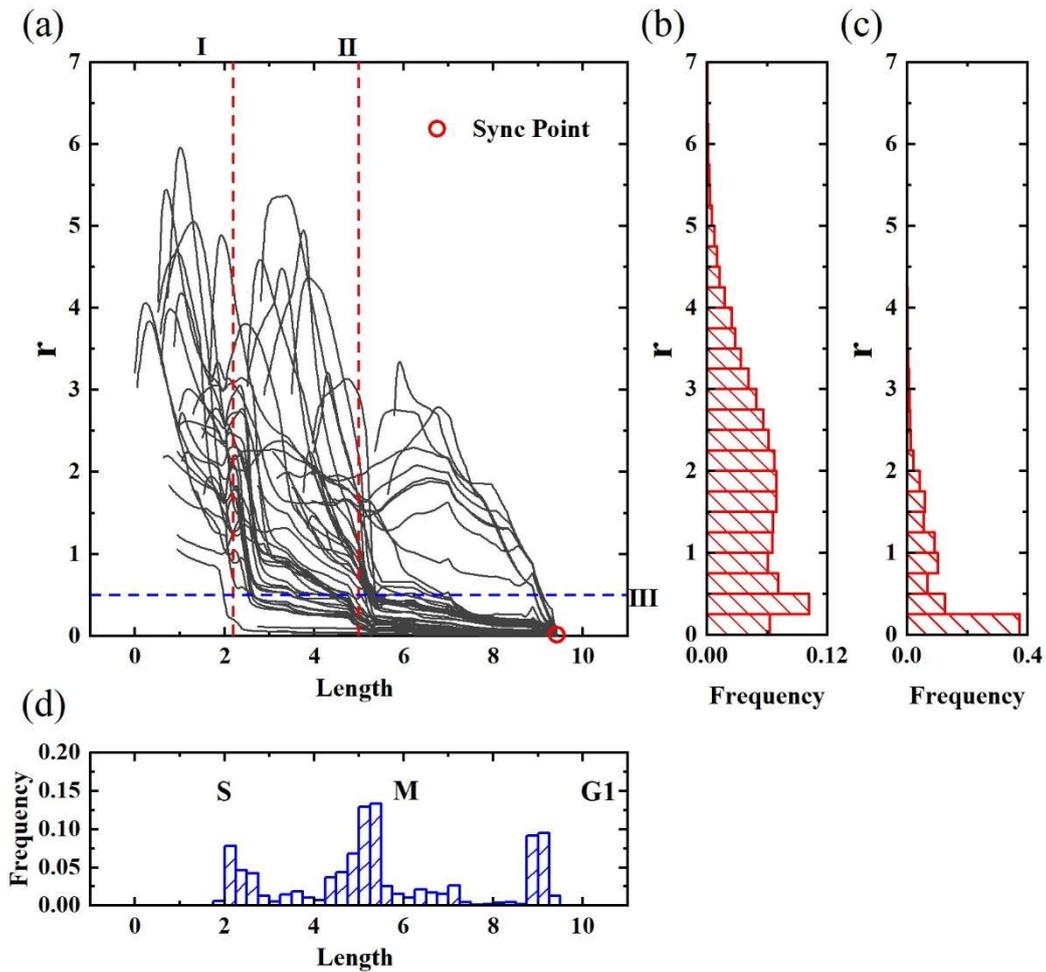

Fig.5 The deviation of $10^5$ random trajectories from the WT process. (a) A sketch of the deviation of random trajectories from the WT process at different WT trajectory distance. The distribution of random trajectories along profile I (b), profile II (c), and profile III (d). The total trajectory counts are 40913 (b), 88971 (c), 82839 (d).

**The new fixed points of ODEs with changing parameters, and the relationship with DNA/Spindle checkpoints**

The structural robustness is an essential feature for an appropriate cellular model (Stelling & Gilles, 2001). It means that the qualitative behaviors of a system are insensitive to parameter perturbations, so the system has ability to "buffer" the variations of environmental conditions. To obtain a global picture of structural stability, we apply the single parameter sensitivity analysis to the G1 attractor and the biological pathway starting from excited G1 (see more details in SI.4).

For G1 attractor, we find that it is sensitive to only 7 parameters in the range of 1/50

~ 50 fold; among them, 5 are in Cln2/SBF feedback loop, while other 2 parameters are relative to Cdh1 (details in Supplementary Table VII). This result shows that the global attractor of resting G1 state is stable to large scale changes of major parameters.

For the biological pathway, one observes that 30 out of 94 parameters can induce bifurcations within the range $\left[\frac{1}{10}, 10\right]$ (distribution in Fig.6a; details in Supplementary Table VIII). Moreover, we also find that the attracting trajectory can still exist even when a new attractor appears (Fig.7b&c): Though some trajectories will be attracted by S phase attractor or early M phase attractor, those out of the capture zone can still follow the attracting trajectory.

We then classify the new attractors, 24 fixed points and 6 limit cycles, from bio-pathway bifurcations within 1/10 ~ 10 fold change (Fig.6b). 24 new fixed points are clustered mainly into two regions (Fig.6c), out of which 18 correspond to cell state with DNA or spindle checkpoint on (illustrated in Fig.7a). This means that even when bifurcation happens, the system is still preferable to stay at states with biological function.

Another biologically meaningful attractor is a limit cycle (LC1) (Fig.6b & Fig.7a). It rises from over-activation of Cln2/SBF loop in G1 phase, triggering the system to pass the threshold of START point without the signal of Cln3.

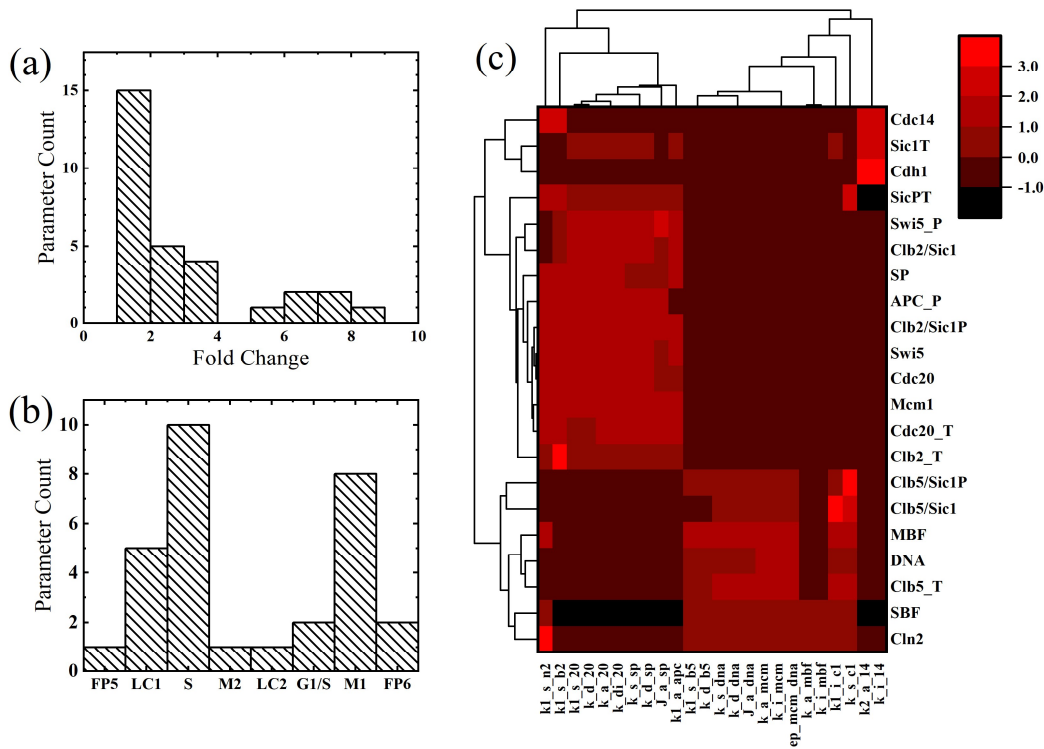

Fig. 6 (a) Distribution of fold change when bifurcation of bio-pathway happens. (b) New attractor types found within [1/10, 10] fold change. G1/S: G1/S arrest; S: S arrest; M1: early M arrest; M2: late M arrest; FP: Fixed point; LC: Limit cycle. (c) Cluster of new fixed points from single parameter perturbation by position in state space (see details in SI.4).

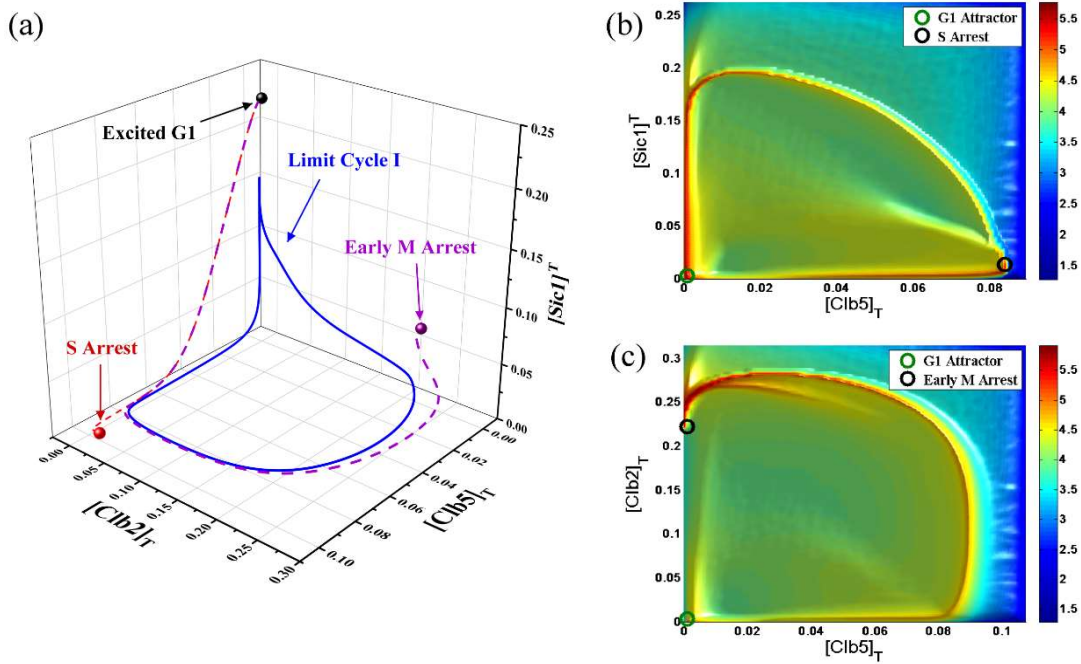

Fig.7 The 2 fixed points and 1 limit cycles as attractors. (a) shows the new attractors which have biological meanings. (b) and (c) are the density maps when S phase attractor or early M phase attractor appears (the circles represent the attractor).

**The saddle-node bifurcations caused by positive feedbacks, and the balance among the coupling S, early M and late M phase modules.**

In the previous subsections, we find that G1 attractor is robust to the changes of major parameters and can be excited by a high level of Cln3. Besides, the attracting WT trajectory will evolve into two checkpoints related fixed points—S arrest and early M arrest—with a high probability when the parameters are changed, which correspond to the turn-on of the checkpoints.

In this section, we analyzed the interactions among G1/S phase module, early M and late M phase modules. In the G1/S phase module, there are positive feedbacks of Cln2 and SBF, Clb5 and MBF, these would deduce the genetic switch with a saddle-node bifurcation in G1/S phase. The concentration of G1 cyclin Cln3 triggered by the cell mass ([Cln3]) works as the key control parameter for the G1/S phase switch. If the

DNA replication checkpoint is turned on and [Cln3] is near the bifurcation point, we could decouple the G1/S phase module from others modules; then we obtained some analysis results about the saddle-node bifurcation and the key control parameters in the G1/S phase (Supplementary Box 1). Similar results can be obtained in the positive feedback of Mcm1/SFF and Clb2 (early M module) and the self-activation of Cdc14 (late M module) (see details in SI.5). Together, we consider respectively [Cln3], $\epsilon_{mcm,dna}$ and $k_{a,20}$ as control parameters to find three saddle-node bifurcations in different modules (Table I), then we chose the three key variables $[Clb5]_T$ in G1/S phase module, $[Clb2]_T$ in early M phase module, and $[Sic1]^T$ in late M phase module to represent the cell-cycle trajectory in Fig.3 and Fig.4.

These parameter sensitivity and bifurcation analysis suggest that the balance among G1/S, early M and late M phase modules ensure the global dynamic robustness in yeast cell-cycle process. Otherwise, the cell-cycle process would arrest at the S phase, early or late M phases or other attractors in Fig.7 (Supplementary Table V).

| Module | Bifurcation | Control parameter | Key variables |
|---|---|---|---|
| G1/S | G1 attractor → S attractor | [Cln3] | $[Clb5]_T$ |
| Early M | S attractor → Early M attractor | $\epsilon_{mcm,dna}$ | $[Clb2]_T$ |
| Late M | Early M attractor → Late M attractor | $k_{a,20}$ | $[Sic1]^T$ |

Table I. Biochemical switches and bifurcations in G1/S, early M and late M module. The variables $[Clb5]_T$, $[Clb2]_T$ and   are chosen to represent the cell-cycle trajectory. All the attractors are listed in Supplementary Table VII & VIII.

**The ideal/perfect yeast cell-cycle model with critical slowing down or ghost effect**

A group of cell-cycle trajectories from different initial states compose the cell-cycle manifold in the state space. When the cell-cycle model utilizes the parameters near the critical points of the above 3 saddle-node bifurcations, where the critical values are $[Cln3]_c = 0.0076$, $(\epsilon_{mcm,dna})_c = 0.24$ and $(k_{a,20})_c = 0.81$, the evolving manifold behaves the critical slowing down or ghost effect with the attractiveness and irreversibility near the saddle-node colliding position. For example, when $\epsilon_{mcm,dna} < (\epsilon_{mcm,dna})_c$, the cell-cycle manifold will be attracted and stay in the S phase attractor in Fig.7. If $\epsilon_{mcm,dna}$ is little large than $(\epsilon_{mcm,dna})_c$, the node point (S phase attractor) and the saddle point collide, the S phase attractor becomes unstable, then the saddle-node remnant will lead a slowly-evolving and converged manifold near the previous S phase attractor. Similarly, when $k_{a,20}$ is little large than $(k_{a,20})_c$, the cell-cycle manifold also converges near the former early M phase attractor in Fig.7.

In Fig.8, the critical slowing down effect is illustrated in the mean transition time maps and density maps. In Fig.8a, using the ideal/perfect cell-cycle model with $\epsilon_{mcm,dna} = 0.32, k_{a,20} = 0.84$, where the system is close to the critical value, the trajectories slow down their paces in S phase (DNA checkpoint) and early M phase (spindle checkpoint) and seem to be bended toward these two regions. However, in Fig.8b, the attracting trajectory by the imperfect cell-cycle model with $\epsilon_{mcm,dna} = 2.0, k_{a,20} = 4.0$, will disperse and no obvious slowing down.

The local perturbation analysis is used to illustrate the role of saddle-node bifurcation in Fig.9 (details in SI.6). Points in a small hypercube around an initial state are taken as perturbed initial states. These perturbation points evolve along the standard trajectory, which starts from excited G1 state, with varying deviations, leading to a change in the diameter of this bunch of trajectories. Therefore, for each point in the standard trajectory, an average radius $r$ can be calculated at the cross section. In Fig.9b&c, we compared the average radius $r$ with different perturbation location and the side length of the hypercube $a$ (WT parameters, $\epsilon_{mcm,dna} = 0.4, k_{a,20} = 1.5$). These results support our former finding that the wild type cell-cycle trajectory attracts the trajectories at specific points in Fig.5, and suggests that the fluctuations are separated by the two checkpoints—the fluctuations in former section will be significantly reduced and have little impact on the latter section. Moreover, we also study the attracting property under different control parameters in Fig.9d: when the system is closer to the critical value $((\epsilon_{mcm,dna})_c = 0.24, (k_{a,20})_c = 0.81)$, the trajectory shows stronger attractiveness.

The ideal/perfect cell-cycle model with ghost effects provides an interesting and unique dynamical perspective for the yeast cell-cycle process. The whole cell-cycle process can be simplified and abstracted to an excitable system composed by three well-coupled saddle-node bifurcations. The cell-cycle trajectory is an attractive trajectory to execute sequential DNA replication and mitosis events with the right order. Because the critical slowing down decouples the sequential events, the cell-cycle trajectory is robust against the fluctuations in both the state and parameter spaces, and the ghost effect offers enough long duration for the execution of each event. Furthermore, the attractive manifold is the simplest mechanism for state checking, and provides the suitable state positions for the checking of completions of both DNA replication and spindle assemble and separation.

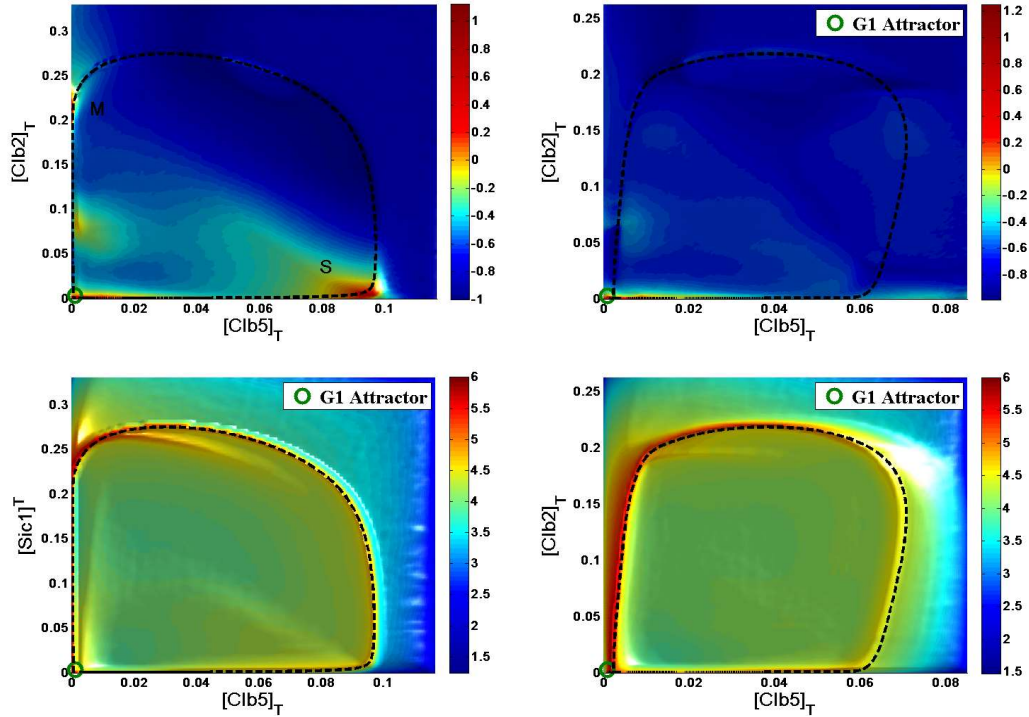

(a) Ideal/perfect cell-cycle model    (b) Imperfect cell-cycle model

Fig.8 The Mean transition time map and trajectory density map of cell-cycle trajectories from $10^6$ random initial states. (a) Ideal/perfect cell-cycle model with $\epsilon_{mcm,dna} = 0.32, k_{a,20} = 0.84$. (b) Imperfect cell-cycle model with $\epsilon_{mcm,dna} = 2.0, k_{a,20} = 4.0$. The phase space is divided into $100 \times 100$ meshes, and the mean transition time or the number of trajectory are calculated in each mesh. The upper two maps record the mean transition time of these trajectories in 2-dimensional space, while the lower record their density. The black dotted lines represent the standard cell cycle trajectory.

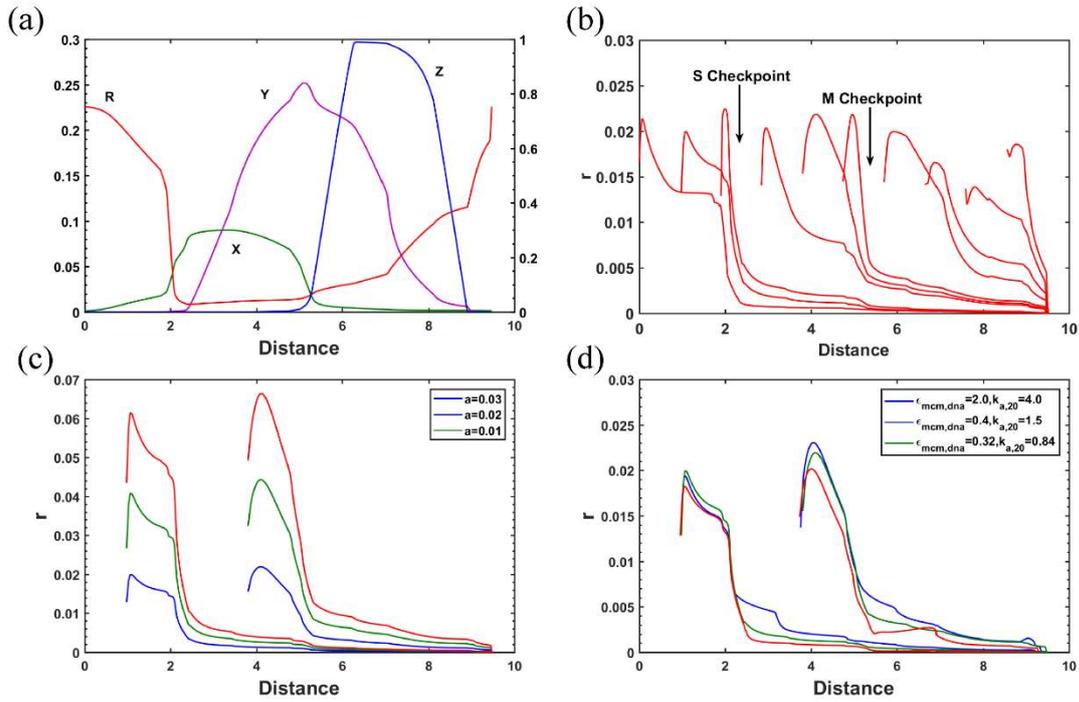

Fig.9 A local dynamic analysis of the wild type cell-cycle trajectories as a function of curve length. (a) plots the evolving trajectory (X: $[Clb5]_T$, Y: $[Clb2]_T$, R: $[Sic1]^T$, left Y-axis; Z: $Cdc14$, right Y-axis); (b, c) plot average radius r with different perturbation location (b) and $a = 0.01, 0.02, 0.03$ (c); (d) plots the attracting property under different $\epsilon_{mcm,dna}$ and $k_{a,20}$ ($a = 0.01$), and the WT trajectory parameter sensitivity analysis shows that the critical values are $(\epsilon_{mcm,dna})_c = 0.24$, $(k_{a,20})_c = 0.81$.

**Modeling the cell-cycle checkpoint mechanism: An "if-then" conditional judgment model**

In the above results of the yeast cell-cycle process by our cell-cycle model, both DNA and spindle checkpoints are assumed to be satisfied automatically, so they are actually turned off during the whole cell-cycle process.

In this section, we discuss the biological mechanism of the cell-cycle checkpoint in budding yeast and the counterpart dynamic model, an "if-then" conditional judgment model. In the real biological yeast cell-cycle process, when DNA replication is incomplete, the DNA replication checkpoint will block mitosis; if the spindle does not assemble or the chromosomes do not properly orient or attach to the spindle, then the spindle checkpoint arrests the mitotic progression. To simulate DNA replication and spindle checkpoint mechanism, we build an "if-then" conditional judgment model, for an example, we set the DNA checkpoint as the following "if-then" conditional judgment role: if $[Clb5]_T$ maintains high level ($> 0.05\ au$) for $T_{DNA} = 15\text{min}$, then turn off the DNA checkpoint by setting [DNA] = 0.4. More details see SI.3. The comparison results of the if-then model (Fig.10) and our continuous ODEs model (Fig.2) shows that both our cell-cycle model can capture the major features of wild type cell-cycle process in budding yeast.

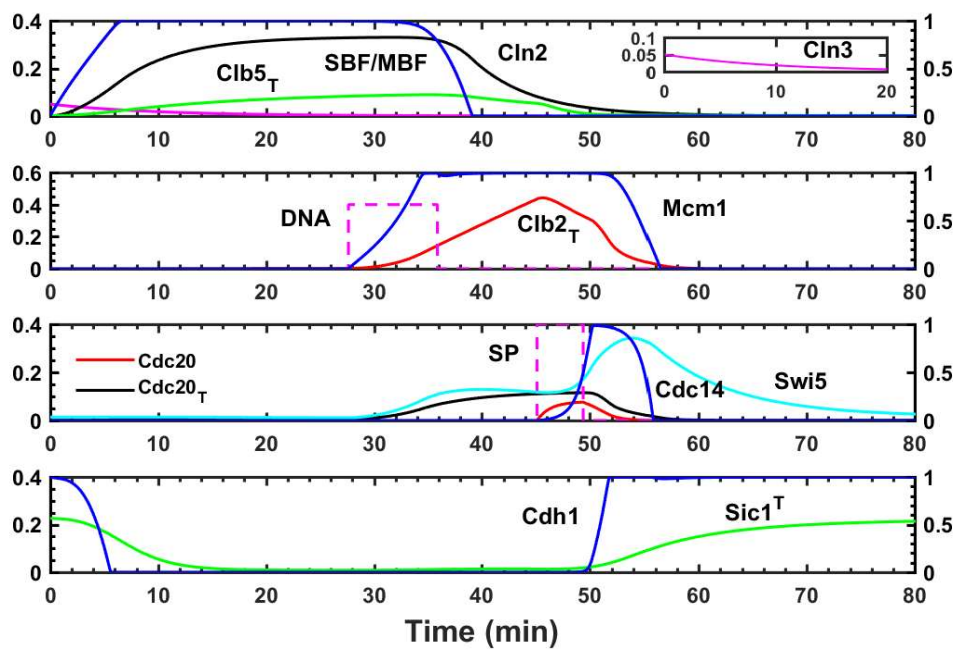

Fig.10 The simulations of wild-type yeast cell-cycle process by the "if-then" checkpoint model. [DNA] and [SP] in the dotted lines are controlled by if-then rules (details in SI.3). The blue line in each panel has a vertical scale from 0 to 1 (right Y-axis), while others are in the range from 0 to 0.4/0.6 (left Y-axis).

**The fundamental structure including repressors that ensures the sequential multiple-event process**

To investigate the global dynamic robustness and the mechanism of sequential events of cell-cycle process in budding yeast, we have constructed three different models, the wild type cell-cycle model, the perfect cell-cycle model with ghost effects, and the "if-then" checkpoint model. Based on the wild type cell-cycle model, the analysis results about the state fluctuation and parameter sensitivity show that: the G1/S phase module, early M and late M modules in the regulatory network, as well as the interactions among the modules play important role in governing the yeast cell-cycle process; the saddle-node bifurcations caused by the positive feedbacks in each module provide the genetic switches for the state transitions. In the perfect cell-cycle model, the system is set to near the critical points of the saddle-node bifurcations, the perfect cell-cycle trajectory exhibits a unique dynamical property for the sequential DNA replication and mitosis events execution, such as the global dynamic robustness and fine-tuned event durations. In the "if-then" checkpoint model, the regulations are more accurate and reliable with a checkpoint pathway to judge the completion of the early event; it, however, is also more complex than the above models.

These results suggest that the whole cell-cycle process is an excitable system with three well-coupled saddle-node bifurcations. Then we obtain the fundamental structure of the yeast cell-cycle regulatory network (Fig.11a) and the essential dynamics of yeast cell-cycle process (Fig.11b). The fundamental cell-cycle network consists of the repressors and G1/S phase module, early M module and late M phase module, where each module contains a positive feedback loop. The repressor proteins are utilized to ensure the sequential events with a certain order. In the beginning of cell-cycle, the repressor is usually in high active level to repress and forbidden the activity of M phase event. After the execution of S phase event, the repressor degrades, and S phase event will trigger M phase event. So the previous event activates the latter one through the checkpoints, while the latter module can also inhibit the previous module. Finally, the last late M phase module turns on the G1 inhibitors to ensure the switch-off of all the modules, this is the G1 state of the cell-cycle process.

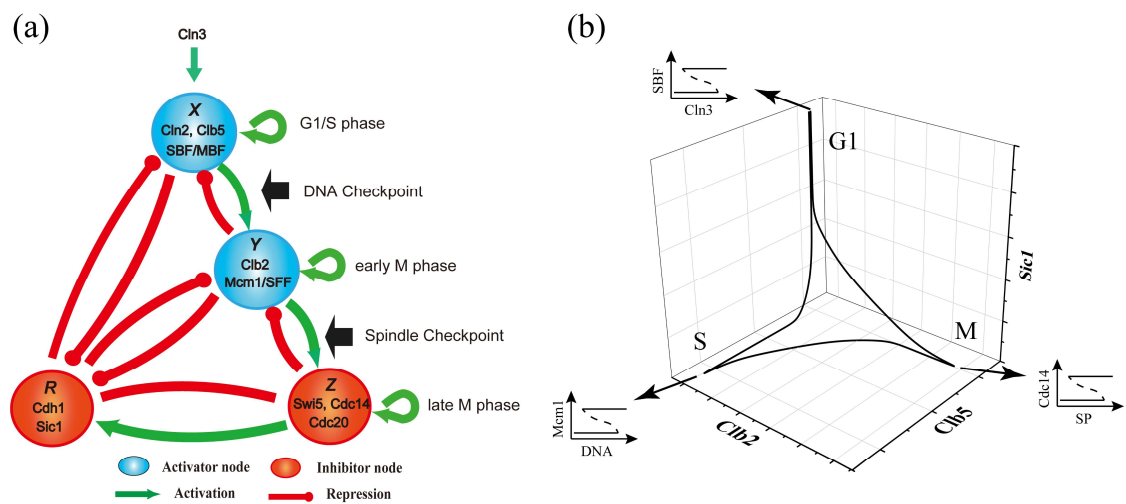

Fig.11 The fundamental structure of yeast cell-cycle network (a) and the essential dynamics of yeast cell-cycle process (b). The yeast cell-cycle process is proposed as an excitable system driven by a sequence of bifurcations with ghost effects; this ensures a globally attractive cell-cycle trajectory that provides a suitable control mechanism for the cell-cycle checkpoints.

Furthermore, we discuss a more abstract and coast-goals picture about the multiple task processes, such as the meiosis processes, differential and development process (Ciliberto & Tyson, 2000), and the flagella forming process in E. coli (Kalir et al., 2001). The repressor protein is also found in flagella formation of E. coli (Aldridge & Hughes, 2002; Kalir et al., 2001).

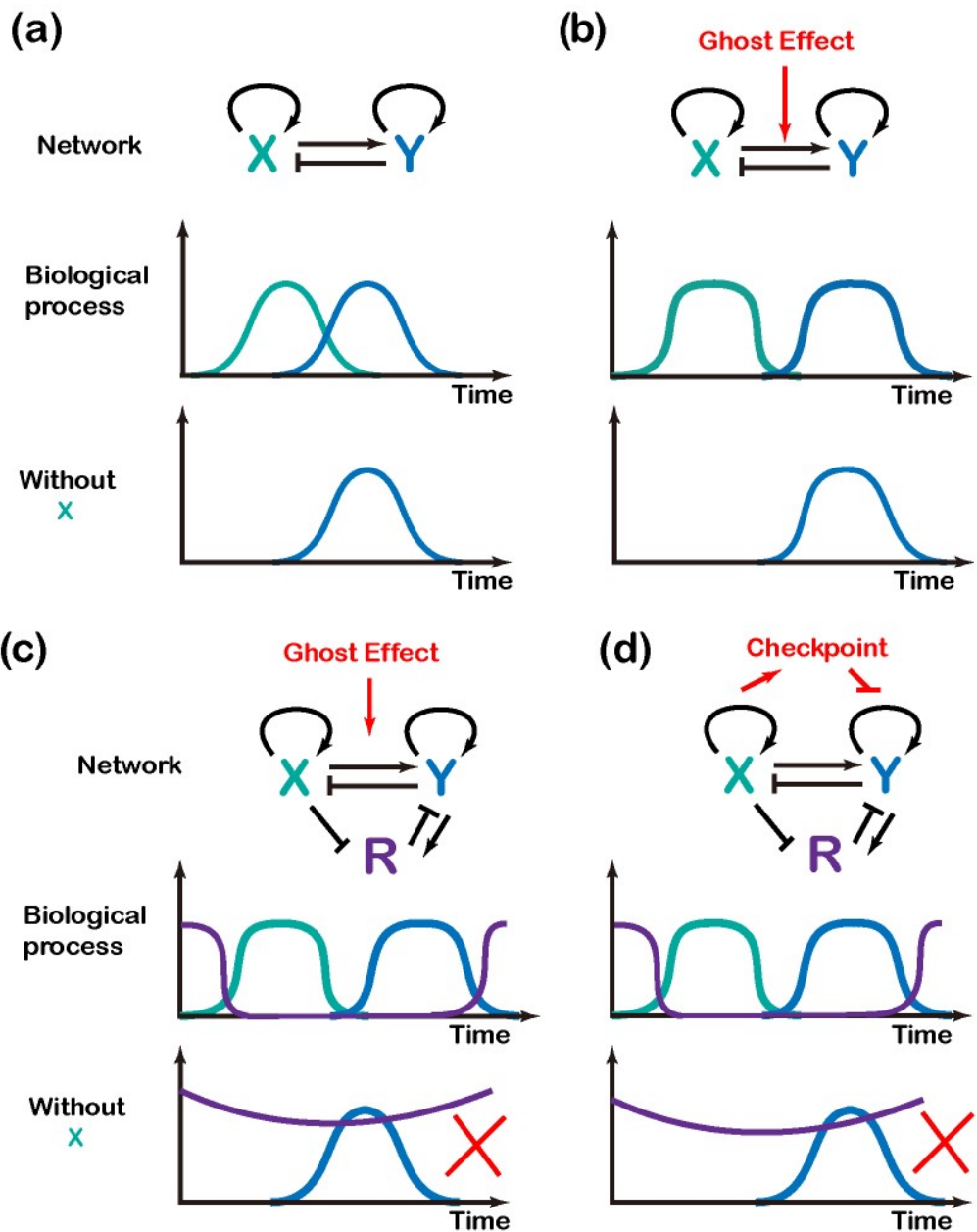

Fig.12 The possible fundamental networks to execute sequential events process. Regulators X and Y respectively control the sequential events A and B. Both the ideal/perfect model containing ghost effects (c) and the "if-then" checkpoint model (d) provide dynamic robustness and sufficient duration for the sequential events process, and forbidden the initiation of event B when event A is not executed.

We assumed a multiple tasks process, event A activates event B, and B is activated only when A is fully activated and finished, where regulators X and Y control A and B respectively. In Fig.12, we illustrate the fundamental regulatory networks and their essential dynamic processes through logical analysis. The simplest structure is the activation of X triggers the activation of Y, the structure X->Y (Fig.12a) and X->Y containing ghost effect (Fig.12b). However, in this situation, the fluctuation of Y may cause the activation of Y without the activation of X. The structure includes the repressor R in Fig.12c, in the beginning of the process R is in an active and high level, thus the activation of Y without the activation of X is forbidden. In Fig.12d, the checkpoint mechanism with repressor R provides double insurance for the order of events A and B. The network with repressor and ghost effect in Fig.12c is similar to the perfect cell-cycle model, this should provide a perfect process for the sequential A and B events with dynamical robustness.

Furthermore, we have utilized the Boolean network model to search the possible 3-node network structures that keep the sequential event order, and find a similar structure as Fig.13c (Li et al., 2013).

After all, we speculate the evolution history of checkpoint mechanism in the eukaryotic cell-cycle control system. A. W. Murray put forward a hypothesis that the evolution of cell-cycle control is from cell mass to the cyclin, and the cyclin with cyclin destruction; he further pointed out that the events of the cell-cycle in early stage would be separated by timing rather than by checkpoints (Murray, 2004). Following the hypothesis, we suppose that, in early stage of cell-cycle process, there did not exist the checkpoint pathway and the checkpoint sensor, kinase and regulator proteins. At that time, the systems were driven by the engine similar to our perfect cell-cycle model with ghost effects, this "ideal or perfect cell-cycle engine" should provide sequential events with the right order and enough event duration, and also offers an attractive manifold for the state checking. This dynamic state checking mechanism should evolve early than the molecular checkpoint pathway. As time went by, in the late evolution stage of cell-cycle, a more reliable and complex checkpoint pathway with sensor, kinase and regulator proteins appeared, this is the analog of the "if-then" conditional judgment model.

## Conclusion

The dynamic robustness and modularity are the important characteristics of the cellular regulatory network for fulfilling complex biological processes. In the eukaryotic cell-cycle process, the cells successively executive DNA replication and mitosis events in order using the checkpoint mechanism. The dynamic regulatory mechanism in cell cycle processes was revealed by the pioneering work of J. Tyson, B. Novak and J. Ferrell et. al (Katherine C. Chen et al., 2004; Csikasz-Nagy, Battogtokh, Chen, Novak, & Tyson, 2006; Ferrell et al., 2011; B Novak, Csikasznagy, Gyorffy, Chen, & Tyson, 1998; Nurse, 2000; Tyson, Chen, & Novak, 2001). The irreversible transitions in the multi-task cell-cycle process are proposed to be regulated by systems-level feedbacks (B. Novak, Tyson, Gyorffy, & Csikasz-Nagy, 2007). The minimal or skeleton models of cell-cycle in budding yeast (Gerard, Tyson, & Novak, 2013) and mammalian cells (Gerard & Goldbeter, 2011) had been established to investigate the temporal order and sequential activation of different phases on cell-cycle processes. And the divergence and convergence manifold along the budding yeast cell-cycle trajectory had been observed and discussed in ODE model (Lovrics et al., 2006) and the Boolean network model (Li et al., 2004).

In this paper, we propose to reveal the global dynamic robustness of cell cycle processes in budding yeast. Based on the previous cell-cycle models and experiments, we constructed a simplified auto-evolving cell-cycle model, where we assumed that DNA replication and spindle assemble and separation processes automatically execute. We find that G1 state is a global stable attractor, and the wild type cell-cycle trajectory is a global attractive trajectory containing several slow evolving parts, which extends and is consistent with our former results in Boolean network model (Li et al., 2004).. In the cell-cycle regulatory network, the G1/S phase module, early M and late phase modules, and the interactions among the modules play important role in governing the yeast cell-cycle process; the saddle-node bifurcations with bistability and hysteresis caused by the positive feedbacks in each module provide genetic switches for the key state transitions in cell-cycle process. Thus, the whole cell-cycle process is an excitable system with three well-coupled saddle-node bifurcations.

If the system is set to near the critical points of the saddle-node bifurcations (ideal/perfect cell-cycle model), the perfect cell-cycle trajectory exhibits a unique dynamical property. The cell-cycle process with critical slowing down or ghost effects is an attractive trajectory to execute DNA replication and mitosis events in order. Because the critical slowing down decouples the sequential events, the cell-cycle trajectory is robust against the fluctuations in both the state and parameter spaces, and the ghost effect offers enough long duration for the execution of each event. Furthermore, the attractive manifold is the simplest mechanism for state checking, and provides the suitable state position for the DNA and spindle checkpoints.

In summary, our results show that the yeast cell-cycle process exhibits as an excitable system containing three well-coupled saddle-node bifurcations to execute DNA replication and mitosis events, and we suggest from the dynamical view a hint from the state checking to the molecule checkpoint pathway in the evolution of eukaryotic cell-cycle processes.

# Acknowledgments

The authors are grateful to Mingyuan Zhong, Mingyang Hu, Xili Liu and Dianjie Li for helpful discussions. The work is supported by NSFC grants nos. 11174011, 11021463, and 91130005 (F.Li), and nos. 11074009 and 10721463 (Q.Ouyang).